# Information Geometric Security Analysis of Differential Phase Shift Quantum Key Distribution Protocol

Laszlo Gyongyosi[*], Sandor Imre

*Quantum Technologies Laboratory*
*Department of Telecommunications*
*Budapest University of Technology and Economics*
2 Magyar tudosok krt,, H-1111, Budapest, Hungary
[*]*gyongyosi@hit.bme.hu*
(Dated: 2010)

**This paper analyzes the information-theoretical security of the Differential Phase Shift (DPS) Quantum Key Distribution (QKD) protocol, using efficient computational information geometric algorithms. The DPS QKD protocol was introduced for practical reasons, since the earlier QKD schemes were too complicated to implement in practice. The DPS QKD protocol can be an integrated part of current network security applications, hence it's practical implementation is much easier with the current optical devices and optical networks. The proposed algorithm could be a very valuable tool to answer the still open questions related to the security bounds of the DPS QKD protocol.**

## 1 Introduction

The DPS QKD scheme was designed to offer a well-implementable and more efficient practical solution with better key generation rates to realize quantum cryptography, than classical QKD approaches. In the DPS quantum cryptography protocol, the sender and the receiver use weak coherent state pulses, and logical bits are encoded in the relative phase of the pulses. The sender encodes every logical bit in two signals, and at the receiver's side, Bob use the two signals to decode the sent logical bit. The relevance of the DPS QKD protocol could have been increased dramatically in practical applications, since the differential phase shift QKD protocol is much more simpler in hardware design than the well known QKD protocols, such as BB84 or the Six-state QKD protocols. On the other side, contrary to it's easy implementation and it's much simpler working mechanism, the DPS QKD's protocol unconditional security is still not proven. Our geometrical analysis shows an efficient algorithmical solution to quantify the secure key generation rate of the DPS QKD protocol, which is still missing from the literature. The possible attacks against the DPS protocol have been studied deeply. This paper analyzes the security of the DPS QKD protocol, using information geometric algorithms, and different quantum cloner models. As the most general attack against the protocol, we analyze coherent attacks, based on two different types of quantum cloner machines.

This paper is organized as follows. First is a short brief on the DPS QKD protocol, and then follows a discussion of the basic elements of computational geometry and quantum information theory. Then we explain the main elements of the present information geometric based approach. Then we show the results on the information-theoretic security analysis of DPS QKD protocol. Finally, we summarize the results.

## 2  DPS QKD Protocol

In today's communication networks, the widespread use of optical fiber and passive optical elements allows to use quantum cryptography in the current standard optical network infrastructure. In the past few years, quantum key distribution schemes have attracted much study. The security of modern cryptographic methods, like asymmetric cryptography, relies heavily on the problem of factoring large integers [3]. In the future, if quantum computers become reality, any information exchange using current classical cryptographic schemes will be immediately insecure [3, 9]. Current classical cryptographic methods are not able to guarantee long-term security. Other cryptographic methods, with absolute security must be applied in the future. Cryptography based on the principles of quantum theory is known as quantum cryptography.
Using current network technology, in order to spread quantum cryptography, interfaces must be implemented that are able to manage together the quantum and classical channels [2, 3, 9]. In practical implementations of QKD protocols, Alice, the sender, uses weak coherent pulses (WCP) instead of a single photon source [47]. As has been shown, WCP based protocols have a security threat, since an eavesdropper can perform a photon number splitting attack against the protocol [1, 2]. These kinds of attacks are based on the fact that some weak coherent pulses contain more than one photon in the same polarization state, which provides information to the eavesdropper without any disturbance. The DPS protocol is robust against such photon number splitting attacks in practice, however a theoretical lower bound on the security of the protocol is still missing from the literature [1, 2]. The working mechanism of the DPS QKD protocol is based on the same idea as the B92 protocol: even two non-orthogonal quantum states are sufficient to perform a secure quantum key distribution. In the DPS protocol, Alice encodes the logical bits in the phase of the pulses. If the phases are modulated by 0, then Alice sends a logical zero, and if the phase between the two pulses is $\pi$, then she encodes a logical one. If the relative phase between two pulses is 0, then Bob will detect 0, and similarly, if the phase between the two pulses is $\pi$, then he will obtain a logical 1.

In the sending process, Alice generates coherent states of the same intensity $\mu$, and from these states she forms a sequence, as follows: [47]

$$\Psi = \ldots \left| e^{i\varphi_{k-1}} \sqrt{\mu} \right\rangle \left| e^{i\varphi_k} \sqrt{\mu} \right\rangle \left| e^{i\varphi_{k+1}} \sqrt{\mu} \right\rangle \ldots = \ldots \left| \psi(k-1) \right\rangle \left| \psi(k) \right\rangle \left| \psi(k+1) \right\rangle \ldots, \quad (1)$$

where the phases can be set at 0 or $\pi$, hence for a logical zero we have $e^{i\varphi_k} = e^{i\varphi_{k+1}}$, and for a logical one, the difference between the two phases is $\pi$, and $e^{i\varphi_k} \neq e^{i\varphi_{k+1}}$. Since the logical bits are encoded in the phases between the signals, the $k$-th signal has relevance in the determination of both the $k$-th and $(k+1)$-th logical bits, hence $\Psi \neq \ldots \left| \psi(k-1) \right\rangle \otimes \left| \psi(k) \right\rangle \otimes \left| \psi(k+1) \right\rangle \ldots$, and this fact increases the complexity of any security analysis [1, 2]. From this viewpoint, the DPS protocol has been analyzed by Takesue and Diamanti *et al.* [18]. In this paper we show that the complexity of the DPS QKD protocol's security analysis can be decreased dramatically, using fast computational information geometric methods.

The security of the DPS QKD protocol lies in the fact that the sender randomly prepares and sends to Bob two non-orthogonal quantum states, similarly to the B92 protocol. The DPS QKD protocol geometrically can be modeled in the same way as the B92 protocol, however its implementation is much easier in practice, since the DPS QKD protocol does not require a bright reference pulse as does the B92 protocol [1, 2]. In practical implementations the DPS scheme can be realized by WCP pulses with an average photon number less than 1, and the sent WCP pulse can be described by

$$\langle \alpha | -\alpha \rangle = e^{-2|\alpha|^2}, \quad (2)$$

where $|\alpha|^2 = \mu \ll 1$ is the average photon number per pulse [1, 2]. The B92 protocol uses the same $0, \pi$ modulation-scheme, however the B92 protocol practical implementation is more complicated than the DPS QKD's scheme [1, 2].

The general setup of the DPS QKD protocol is illustrated in Figure 1. The time difference between the pulses is known at the receiver's device.

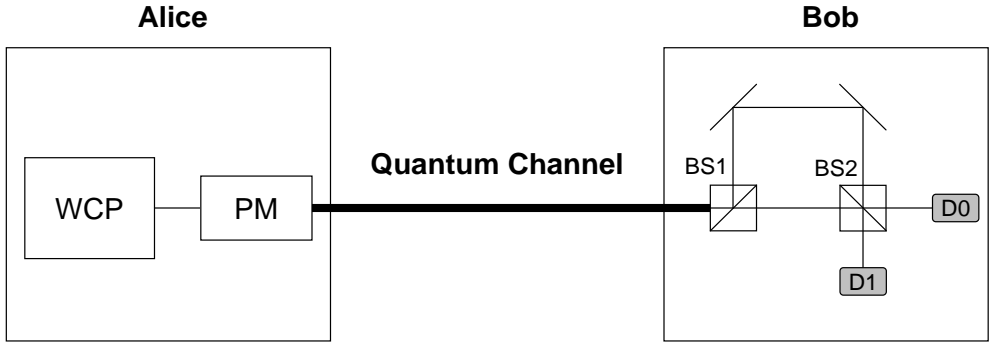

**Figure 1.** A schematic view of DPS QKD protocol. Alice uses Weak Coherent Pulses (WCP) and a Phase Modulator (PM) to generate the signals. Bob decodes them with Beam Splitters (BS1, BS2) and photon detectors (D0, D1).

The optimal secure key rate of the protocol has been guaranteed only for individual attacks where the eavesdropper acts on the photons individually [1, 2]. In our geometrical security analysis, we will analyze the most general collective attacker model, since the security of DSP QKD protocol against this general attack still remains an open question. In this attacker model, we analyze only the cloned photons from the given pulse, and we will give an approximation on the information obtainable by the eavesdropper. Our analysis focuses on the eavesdropper's information about the given key, and the uses the eavesdropper's cloned quantum states. In the experimental realization of DPS QKD protocol, the signal consists of a weak coherent state and a strong phase reference. The relative optical phase between weak coherent state and reference pulse is either 0 or $\pi$, and these kinds of signals were already used in classical QKD schemes. In our attacker model, for simplicity we model these signals as two non-orthogonal quantum states, and we will analyze the still open questions related to the lower bounds on eavesdropper's obtainable information. As has been shown by Inoue [1] and Honjo [2], the photon number splitting attacks can not be realized only with zero-error, and several other attacks have been studied by Curty, Tamaki, and Moroder [19], Waks, Takesue, and Yamamoto [20], Branciard et al. [21], Curty et al. [22], Tsurumaru et al. [23], Branciard, Gisin, and Scarani [24], and Gomez-Sousa and Curty [25]. In this paper, we will use the results of Inoue et al. [1] and Honjo et al. [2]. We will analyze the most common attacker model against DPS protocol, to analyze geometrically its information-theoretical security.

## 2.1 Computational Geometry in QKD Analysis

Many challenging hard algorithmic problems can be studied with computational geometry and, at present, there exist many geometric algorithms that offer an efficient and well implementable solution for hard computational problems. Computational Geometry was originally focused on the construction of efficient algorithms and it provides a very valuable and efficient tool for computing hard tasks [13]. In many cases, the traditional linear programming methods are not very efficient. The computation of the convex hull between quantum states cannot be computed efficiently by linear programming, however the methods of computational geometry are better at solving these kinds of hard problems [5, 6, 7, 8, 15]. Computational Geometry uses the results of classical geometry and the power of computing. In Figure 2, we illustrate the logical structure of the analysis and the cooperation of classical and quantum systems. To this day, the most efficient classical algorithms for this purpose are computational geometric methods. We use these classical computational geometric tools to analyze the security of DPS QKD protocol.

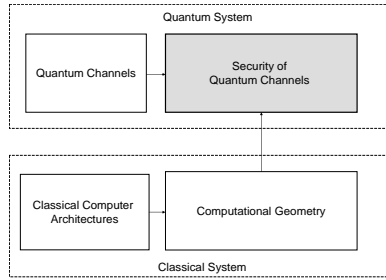

**Figure 2.** Logical structure of our analysis. We use current classical architectures to analyze the properties of quantum channels.

In this paper, we use the methods of computational geometry to analyze the security of quantum channels, however we use quantum information as a distance measure instead of classical geometric distances. Unlike ordinary geometric distances, a quantum informational distance is not a metric and it is not symmetric, hence this pseudo-distance features as a measure of informational distance. This paper combines the results of information geometry and the fast methods of computational geometry. Using the quantum informational distance as a distance measure, we analyze the privacy of eavesdropped quantum channels [17].

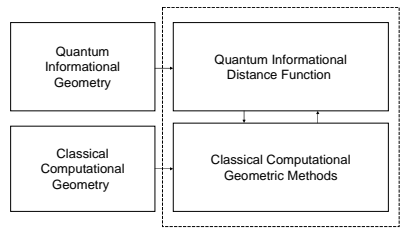

**Figure 3.** Quantum information as distance measure in classical computational geometric methods.

From the combination of the quantum informational distance function and classical computational geometric methods, the properties of quantum channels and quantum states in quantum space can be analyzed as geometrical objects in geometrical space. The application of information geometry in the quantum space has been studied by Nielsen and Nock [31-36], Kato et al. [46], Hayashi et al. [38] and Cortese [40-41]. The properties of quantum informational distance and quantum relative entropy function have been analyzed by Ohya and Petz [42-43], Petz [44-45] and Hayashi [39].

## 2.2 Preliminaries

The security of QKD schemes relies on the no-cloning theorem [3, 4, 10, 11, 12]. Contrary to classical information, in a quantum communication system, quantum information cannot be copied perfectly. If Alice sends a number of photons $|\psi_1\rangle, |\psi_2\rangle, \ldots, |\psi_N\rangle$ through the quantum channel, an eavesdropper is not interested in copying an arbitrary state, only the possible polarization states of the attacked QKD scheme. The unknown states cannot be cloned perfectly, the cloning process of quantum states is possible only if the information being cloned is classical, hence the quantum states are all orthogonal. The polarization states in the QKD protocols are not all orthogonal states, which makes it impossible for an eavesdropper to copy the sent quantum states [3, 9]. Our goal is to measure the level of quantum cloning activity on the quantum channel, using fast computational geometric methods. We measure the information-theoretical impacts of quantum cloning activity in the quantum channel. Alice's side is modeled by a random variable $X = \{p_i = P(x_i)\}, i = 1, \ldots N$. Bob's side can be modeled by another random variable $Y$. The Shannon entropy for the discrete random variable $X$ is denoted by $H(X)$, which can be defined as $H(X) = -\sum_{i=1}^{N} p_i \log(p_i)$, for conditional random variables, the probability of random variable $X$ given $Y$ is denoted by $p(X|Y)$. Alice sends a random variable to Bob, who produces an output signal with a given probability. We analyze in a geometrical way the effects of Eve's quantum cloner on Bob's received quantum state. Eve's cloner in the quantum channel increases the uncertainty in $X$, given Bob's output $Y$. The information-theoretical noise of Eve's quantum cloner increases the conditional Shannon entropy $H(X|Y)$, where $H(X|Y) = -\sum_{i=1}^{N_X} \sum_{j=1}^{N_Y} p(x_i, y_j) \log p(x_i|y_j)$. Our geometrical security analysis is focused on the cloned mixed quantum state received by Bob.

This paper analyzes the phase-covariant cloner and the universal quantum cloner (UCM) model, since both quantum cloners can be applied in attacks against DPS QKD protocol [10, 11, 12]. In the general attacker model, Alice's pure state is denoted by $\rho_A$, Eve's cloner is modeled by an affine map $\mathcal{L}$ and Bob's mixed input state is denoted by $\mathcal{L}(\rho_A) = \sigma_B$. In our calculations, we can use the fact that for random variables $X$ and $Y$, $H(X,Y) = H(X) + H(Y|X)$, where $H(X)$, $H(X,Y)$ and $H(Y|X)$ are defined in terms of probability distributions $p(x), p(x,y)$ and $p(y|x)$. We measure in a geometrical representation the information which can be transmitted in the presence of an eavesdropper on the quantum channel.

In Figure 4, we illustrate Eve's quantum cloner on the quantum channel. Alice's pure state is denoted by $\rho_A$, the eavesdropper's quantum cloner transformation is denoted by $\mathcal{L}$. The mixed state received by Bob is represented by $\sigma_B$. In our information geometric security analysis, we would like to compute the information obtainable by the eavesdropper. The informational-theoretic security of the DPS QKD protocol is analyzed by the radius of the eavesdropper's smallest enclosing quantum informational ball.

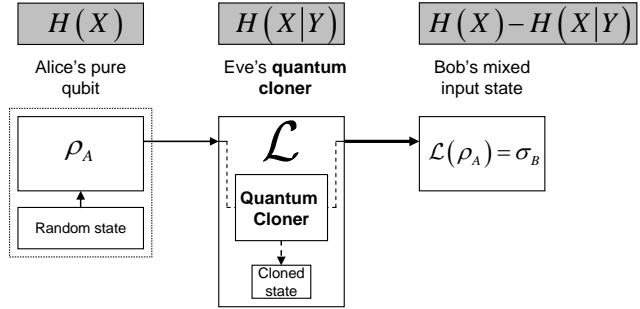

**Figure 4.** The analyzed attacker model and entropies.

In general, for a private communication channel, we seek to maximize $H(X)$ and minimize $H(X|Y)$ in order to maximize the radius of the smallest enclosing ball, which describes the maximal transmittable information from Alice to Bob in the attacked quantum channel. Classically, this information can be expressed as $H(X) - H(X|Y)$.

In our information-theoretic security analysis, we would like to measure the information which can be obtained by the eavesdropper, using advanced collective attacker model. As we will see, this quantity can be described by the radius of the smallest enclosing quantum informational ball of the eavesdropper's, using the Holevo quantity. Geometrically, the presence of an eavesdropper causes a detectable mapping to change from a noiseless one-to-one relationship to a stochastic map. If there is no cloning activity on the channel, then $H(X|Y) = 0$, and the radius of the smallest enclosing quantum informational ball on Bob's side will be maximal.

## 2.3 Measuring Quantum Informational Distances between Quantum States

A quantum state can be described by its density matrix $\rho \in \mathbb{C}^{d \times d}$, which is a $d \times d$ matrix, where $d$ is the level of the given quantum system. For an $n$ qubit system, the level of the quantum system is $d = 2^n$. We use the fact that particle state distributions can be analyzed probabilistically by means of density matrices. A two-level quantum system can be defined by its density matrices in the following way:

$$\rho = \frac{1}{2}\begin{pmatrix} 1+z & x-iy \\ x+iy & 1-z \end{pmatrix}, \quad x^2 + y^2 + z^2 \leq 1, \tag{3}$$

where $i$ denotes the complex imaginary $i^2 = -1$. The density matrix $\rho = \rho(x, y, z)$ can be identified with a *point* $(x, y, z)$ in 3-dimensional space, and a ball $\mathcal{B}$ formed by such points $\mathcal{B} = \{(x, y, z) | x^2 + y^2 + z^2 \leq 1\}$, is called a Bloch-ball. The eigenvalues $\lambda_1, \lambda_2$ of $\rho(x, y, z)$ are given by $\left(1 \pm \sqrt{x^2 + y^2 + z^2}\right)/2$, the eigenvalue decomposition $\rho$ is $\rho = \sum_i \lambda_i E_i$, where $E_i E_j$ is $E_i$ for $i = j$ and 0 for $i \neq j$. For a mixed state $\rho(x, y, z)$, $\log \rho$ defined by $\log \rho = \sum_i (\log \lambda_i) E_i$. In quantum cryptography the encoded pure quantum states are sent through a quantum communication channel. Using the Bloch sphere representation, the quantum state $\rho$ can be given as a three-dimensional point $\rho = (x, y, z)$ in $\mathbb{R}^3$ and it can be represented in spherical coordinates $\rho = (r, \theta, \varphi)$, where $r$ is the radius of the quantum state to the origin, $\theta$ and $\varphi$ represents the latitude and longitude rotation angles.

The classical Shannon-entropy of a discrete $d$-dimensional distribution $p$ is given by $H(p) = \sum_{i=1}^{d} p_i \log \frac{1}{p_i} = -\sum_{i=1}^{d} p_i \log p_i$. The *von Neumann* entropy $\mathsf{S}(\rho)$ of quantum states is a generalization of the classical Shannon entropy to density matrices [6, 7]. The entropy of quantum states is given by $\mathsf{S}(\rho) = -Tr(\rho \log \rho)$. The quantum entropy $\mathsf{S}(\rho)$ is equal to the Shannon entropy for the eigenvalue distribution $\mathsf{S}(\rho) = \mathsf{S}(\lambda) = -\sum_{i=1}^{d} \lambda_i \log \lambda_i$, where $d$ is the level of the quantum system. The relative entropy in classical systems is a measure that quantifies how close a probability distribution $p$ is to a model or candidate probability distribution $q$ [4, 6]. For $p$ and $q$ probability distributions, the relative entropy is given by $D(p\|q) = \sum_i p_i \log_2 \frac{p_i}{q_i}$, while the relative entropy between quantum states is measured by

$$D(\rho\|\sigma) = Tr[\rho(\log \rho - \log \sigma)]. \quad (4)$$

To compute the radius $r^*$ of the smallest informational ball of quantum states and the entropies between the cloned quantum states, instead of classical Shannon entropy, we can use von Neumann entropy and quantum relative entropy.

## 2.4 Security Problem in Quantum Cryptography

In quantum cryptography, the best eavesdropping attacks use quantum cloning machines [10, 11, 12]. However, an eavesdropper cannot measure the state $|\psi\rangle$ of a single quantum bit, since the result of her measurement is one of the single eigenstates of the quantum system. The measured eigenstate gives only very poor information to the eavesdropper about the original state $|\psi\rangle$ [3, 9]. The eavesdropper's cloning transformation is a trace-preserving and completely positive map and it can be described as $\{\mathcal{L},|Q\rangle\}$, where $|Q\rangle$ is the eavesdropper's ancilla state. The process of cloning pure states can be generalized as

$$|\psi\rangle_a \otimes |\Sigma\rangle_b \otimes |Q\rangle_x \to |\Psi\rangle_{abc}, \quad (5)$$

where $|\psi\rangle$ is the state in Hilbert space to be copied, $|\Sigma\rangle$ is a reference state and $|Q\rangle$ is the ancilla state [10]. As Wooters and Zurek have shown, an unknown quantum state $|\psi\rangle = \alpha|0\rangle + \beta|1\rangle$ cannot be cloned perfectly [9], however it was later shown that an unknown quantum state can be cloned approximately [11, 12]. A cloning machine is called symmetric if at the output all the clones have the same fidelity, and asymmetric if the clones have different fidelities [11, 12]. The effect of the eavesdropper's quantum cloner simply shrinks the Bloch-ball with probability $p$. The general model of Eve's cloning machine is shown in Figure 5 [6, 11, 12, 13].

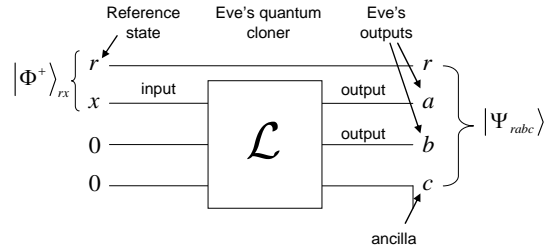

**Figure 5.** General model of Eve's quantum cloner.

The input qubit state is denoted by $x$, which is initially in an entangled state with a reference qubit $r$, denoted by the Bell state $\left|\Phi^+\right\rangle_{rx}$. After the cloning transformation, the overall system consists of the three outputs and the reference quantum state, thus the output state $\left|\Psi_{rabc}\right\rangle$ can be written as a superposition of double Bell states [9]:

$$\left|\Psi_{ra,bc}\right\rangle = v\left|\Phi^+\right\rangle\left\langle\Phi^+\right| + z\left|\Phi^-\right\rangle\left\langle\Phi^-\right| + x\left|\Psi^+\right\rangle\left\langle\Psi^+\right| + y\left|\Psi^-\right\rangle\left\langle\Psi^-\right|, \qquad (6)$$

where $x, y, z$ and $v$ are complex amplitudes with $|x|^2 + |y|^2 + |z|^2 + |v|^2 = 1$. The qubit pairs $ra$ and $bc$ are Bell mixtures with $|x|^2 = p_x$, $|y|^2 = p_y$, $|z|^2 = p_z$ and $|v|^2 = 1 - p$. Equation $v = x + y + z$, describes a three-dimensional surface in the space, where each point $(x, y, z)$ represents parameters $x^2 = p_x$, $y^2 = p_y$ and $z^2 = p_z$. This surface is an oblate *ellipsoid* $\mathcal{E}$ [3, 10, 11, 12] and we denote the coordinates of the ellipsoid $\mathcal{E}$ by $(x_\mathcal{E}, y_\mathcal{E}, z_\mathcal{E})$. The ellipsoid $\mathcal{E}$ has polar radius $x_\mathcal{E} = \frac{1}{2}$, while the equatorial radius is $z_\mathcal{E} = 1$ [10, 15]. In Figure 6, we have illustrated the effects of phase-covariant and universal quantum cloners that have been analyzed.

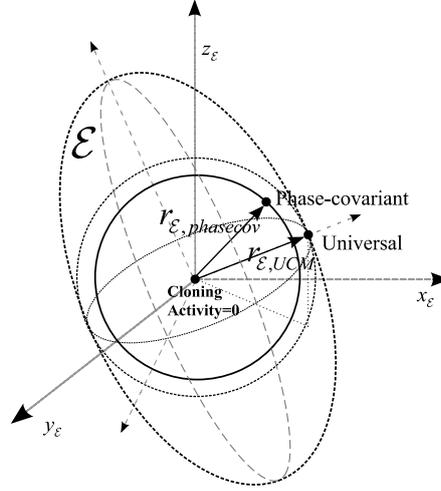

**Figure 6.** Comparison of UCM and phase-covariant based attack in ellipsoidal representation.

The radii which describe the shrinking of the cloned state are denoted by $r_{\mathcal{E}, phasecov}$ and $r_{\mathcal{E}, UCM}$ in the three-dimensional ellipsoidal representation.

# 3 Cloning Machine-Based Attacks in Quantum Cryptography

Our security analysis is based on the DPS quantum cryptography protocol. In this setting, we model the attacks against the DPS QKD protocol as follows: Alice sends a qubit $|\psi\rangle$ to Bob and Eve clones the sent qubit using her ancilla qubit $|E\rangle$. In the cloning process, Eve's ancilla state $|E\rangle$ interacts with the sent qubit $|\psi\rangle$ and the unknown state is forwarded to Bob who makes his standard measurement. To realize a coherent attack, it can be assumed that Eve can store her qubit, and she can make a post measurement on the collected qubits [3]. If Eve uses a phase-covariant cloning machine, then she clones only equatorial states $|\psi\rangle = \frac{1}{\sqrt{2}}\big(|0\rangle + e^{i\varphi}|1\rangle\big)$. In that case, if she uses a universal cloner, then she clones all the states, hence if the value of the parameter $F_{Eve}$ is independent of the input quantum state $|\psi\rangle$. The quantum cloning transformation is optimal [10, 12], if $\eta = \frac{2}{3}$, and hence the maximum fidelity of optimal universal cloning is $F_{Eve} = \frac{5}{6}$, and the maximum radius is $r_{Eve}^{UCM} = \frac{2}{3}$. The quantum information-theoretical radius can be defined as $r_{Eve}^{*UCM} = 1 - \mathsf{S}\big(r_{Eve}^{UCM}\big)$, where $\mathsf{S}$ is the von Neumann entropy of the corresponding quantum state with radius $r_{Eve}^{UCM}$. In general, the universal cloning machine output state can be given as [10, 11, 12]

$$\rho^{(out)} = F_{Eve}|\psi\rangle_a\langle\psi| + \big(1 - F_{Eve}\big)|\psi_\perp\rangle_a\langle\psi_\perp|. \tag{7}$$

We note, that universal cloning also has direct application to eavesdropping strategies in *Six-state* quantum cryptography. Here, we study the information geometric properties of this cloner from the viewpoint of DPS QKD protocol. The optimal fidelity of 1 to 2 phase-covariant cloning transformation is given by $F_{1\to 2}^{pha\,sec\,ov.} = \frac{1}{2} + \sqrt{\frac{1}{8}} \approx 0.8535$ [10, 11, 12]. If Eve has a phase-covariant quantum cloner, then the maximum value of her radius is $r_{Eve}^{pha\,sec\,ov.} = 2\sqrt{\frac{1}{8}}$. The quantum information-theoretical radius $r_{Eve}^{*pha\,sec\,ov.}$ of the phase-covariant cloner can be defined as $r_{Eve}^{*pha\,sec\,ov.} = 1 - \mathsf{S}\big(r_{Eve}^{phasecov.}\big)$. The phase-covariant quantum cloning transformation produces two copies of the equatorial qubit with optimal fidelity.

## 3.1 Proposed Model for Quantum Cloning-Based Eavesdropping Detection

The information-theoretical impacts of the eavesdropper's cloning machine are measured by the radius $r^*$ of the smallest enclosing quantum informational ball. We use the Delaunay tessellation, because it is the fastest known tool for seeking the center of the smallest enclosing ball of points [7, 16]. As the first part of our theorem, for a secure quantum channel, the radius $r^*$ of the smallest enclosing quantum information ball of mixed states has to be greater than $r^*_{Eve}$, thus $r^* > r^*_{Eve}$. As the second part, for an insecure quantum channel, the radius $r^*$ is smaller than or equal to $r^*_{Eve}$, thus $r^* \leq r^*_{Eve}$. In Figure 7, we show a geometrical interpretation of our model for a secure and for an attacked quantum channel [3, 17].

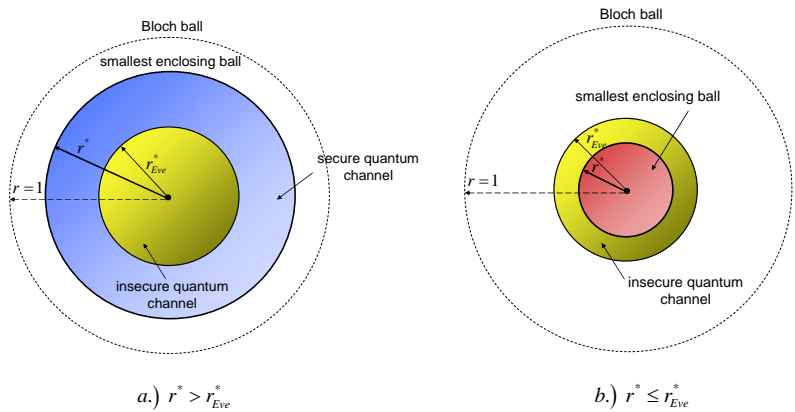

**Figure 7.** Radius of the smallest enclosing information ball for secure and attacked quantum communication.

The information-theoretical effect of the eavesdropper's cloning machine is described by the radius $r^*$ of the smallest enclosing quantum informational ball. The quantum information-theoretical radius $r^*$ is equal to the maximum quantum informational distance from the center and can be expressed as:

$$r^* = \min_{\sigma} \max_{\rho} D\big(\mathcal{L}(\rho) \big\| \mathcal{L}(\sigma)\big). \tag{8}$$

In our geometrical approach, we compute the smallest enclosing information ball by Delaunay tessellation, which is the fastest known tool to seek the center of a smallest enclosing ball of points [7, 16]. For UCM and phase-covariant cloning, the connection between information-theoretical radius $r^*$ and Bloch vector $r_{Bloch}$ can be defined as:

$$r^* = 1 - \mathsf{S}(r_{Bloch}). \tag{9}$$

The information-theoretical radius of UCM and phase-covariant cloners are denoted by $r^*_{UCM}$ and $r^*_{phasecov.}$.

## 3.2 Attacks against the DPS QKD Protocol

To analyze the possible effects of an eavesdropper, we will use two kinds of quantum cloners, the most general universal quantum cloner and the phase covariant cloner. Other possible attacks against the protocol, such as sequential attacks, unambiguous state discrimination attacks or minimum error discrimination methods have been analyzed, and upper bounds have been obtained [1, 2].

### 3.2.1 Collective Attack

This paper analyzes the most general attack against the DPS QKD protocol, the collective attacks. The security bounds for this type of attack were analyzed by Biham and Mor [26], and they have concluded the same bound holds for the protocol. In this attacker model, the eavesdropper tries to clone each of the quantum states sent by Alice, following an independent cloning strategy. The eavesdropper can change her strategy in a probabilistic way, hence in practical QKD applications, Eve can stop her cloning activity for a while, and then later, she can attack again. By changing the used strategies, she can decrease the probability of detection of her activity in the quantum channel.

In a collective attack, an eavesdropper can use a quantum memory to store her quantum states, and she can delay the whole measurement process. She can collect the required information from the steps of key agreement between Alice and Bob, which can be used to choose the best measurement strategy on the collected quantum states.

As has been shown by Devetak and Winter [27], the generic security bound for an collective attack can be given by the Csiszár–Körner bound [28] for one-way postprocessing as $I(A:B) - \min(I_{AE}, I_{BE})$ with $I_{AE} = \max_{Eve} \chi(A:E)$, where $I_E$ is the eavesdropper's information about the raw key of Alice and Bob, and $I(A:B)$ can be expressed as

$$H(A) + H(B) - H(AB). \tag{10}$$

To compute the $\chi$ *Holevo quantity*, we will use our geometrical approach, with quantum relative entropy as distance measure. Geometrical, this quantity can be expressed as the radius of the smallest enclosing quantum informational ball, hence

$$\begin{aligned} r^* &= I(A:E) = \mathsf{S}(E) - \mathsf{S}(E|A) = \\ \max \chi(A:E) &= \max\left(\mathsf{S}(\rho_E) - \sum_a p(a)\mathsf{S}(\rho_{E|a})\right), \end{aligned} \tag{11}$$

where $\chi$ is the Holevo quantity, $a$ is Alice's output with probability distribution $p(a)$, and $\rho_{E|a}$ is Eve's ancilla and $\rho_E = \sum_a p(a) \rho_{E|a}$ is the partial state of the eavesdropper. We note, that the same equation can be applied between Alice and Bob, when Bob is also able to store the quantum states.

In our geometrical approach, we will use the radius of the smallest enclosing quantum informational ball to approximate the eavesdropper's maximal obtainable information. The size of the smallest quantum informational ball depends only on the properties of the eavesdropper's states $\rho_{E|a}$, and it is independent of the chosen measurement strategy, hence our analysis can be extended to not just a very specific examples of set of cloned quantum states.

We note that the eavesdropper's most general strategy can include many possible variations which cannot be parameterized efficiently. However the security bounds for general or coherent attacks are the same as for collective attacks, hence our geometrical approach can be used to analyze both collective and coherent attacks.

*3.2.2 Collective Beam-Splitting Attack against the DPS QKD*

As has been shown by Branciard, Gisin, and Scarani [24], the simplest realization of a collective attack against the DPS QKD protocol is the beam-splitting attack, hence in this paper we use this type of attack to describe the informational-theoretic security of the DPS QKD protocol.

To describe the coherent beam-splitting attack, we model Alice's sent states as a sequence of coherent states $\otimes_i |\psi(i)\rangle$, where each $\psi(i)$ is chosen from the set $\{+\psi, -\psi\}$, and the logical value of the bit is 0 if $\psi(i-1) = \psi(i)$, and 1 if $\psi(i-1) = -\psi(i)$. In the collective beam-splitting attack, the eavesdropper uses a beamsplitter to get a fraction of the signal. The remaining fraction of the signal, denoted by $\tau$, is sent directly to Bob, hence Bob will receive the state $\otimes_i |\psi(i)\sqrt{\tau}\rangle$, and similarly, the eavesdropper's state can be described as $\otimes_i |\psi(i)\sqrt{1-\tau}\rangle$. The eavesdropper's information can be given by using the von Neumann entropy, as

$$I_E^{DPS} = \mathsf{S}(\rho_E) - \frac{1}{2}\mathsf{S}(\rho_{E|0}) - \frac{1}{2}\mathsf{S}(\rho_{E|1}), \tag{12}$$

where it is assumed that the probability of each logical bit value is equal, hence

$$\rho_E = \frac{1}{2}\rho_{E|0} + \frac{1}{2}\rho_{E|1}. \tag{13}$$

Using the coding scheme $\psi(i-1) = \psi(i)$ for 0, and $\psi(i-1) = -\psi(i)$ for 1, the state of $\rho_{E|0}$ and $\rho_{E|1}$ can be expressed as

$$\rho_{E|0} = \frac{1}{2}P_{+\psi_E,+\psi_E} + \frac{1}{2}P_{-\psi_E,-\psi_E} \text{ and } \rho_{E|1} = \frac{1}{2}P_{+\psi_E,-\psi_E} + \frac{1}{2}P_{-\psi_E,+\psi_E} \quad (14)$$

where $\psi_E = \psi\sqrt{1-\tau}$ and $P_{\psi_E}$ is the projector. Using $\psi_E$, we can introduce a new parameter,

$$\gamma = e^{-|\psi_E|^2} = e^{-\mu(1-\tau)}, \quad (15)$$

where $\mu$ is the intensity of the sent weak coherent pulse. Using this parameter, the inner product between $|\langle +\psi_E | -\psi_E \rangle| = \gamma^2$, for given $\mu$ intensity, the eavesdropper's information can be expressed as

$$I_E^{DPS}(\mu) = 2H\left[\frac{(1-\gamma^2)}{2}\right] - H\left[\frac{(1-\gamma^4)}{2}\right] = \\ 2H\left[\frac{(1-|\langle +\psi_E | -\psi_E \rangle|)}{2}\right] - H\left[\frac{(1-|\langle +\psi_E | -\psi_E \rangle|^2)}{2}\right], \quad (16)$$

where $H$ is the binary entropy function.

Using our geometrical approach, the connection between the practically achievable secret key rate $K$ of the protocol and the radius $r^*$ of the smallest enclosing quantum informational ball of the eavesdropper can be given by:

$$K(\mu) = \left[I(A:E) - r^*\right]R = \left[I(A:B) - \max_{Eve}\left(\mathsf{S}(\rho_E) - \sum_a p(a)\mathsf{S}(\rho_{E|a})\right)\right]R = \\ v\left(1 - e^{-\mu\tau}\right)\left(1 - I_E^{DPS}(\mu)\right), \quad (17)$$

where $R$ is the raw key rate and $v$ is the repetition rate.

In the numerical analysis section of this paper, we will show the results of our information-theoretic security analysis of QKD protocol. We will describe the maximal obtainable information of the eavesdropper by the radius of the smallest enclosing quantum informational ball.

### 3.3 Geometrical Modeling of DPS QKD Protocol

In phase-coding QKD schemes, a signal consists of a superposition of two time-separated pulses. These methods, instead of polarization encoding, encode the information in the relative phase between two pulses. However, the polarization and phase encoding schemes are equivalent mathematically [1, 2], hence in our information geometrical security analysis, we can use the Bloch-ball representation to study the security of the protocol. We can use the following translation between the basis states $|0\rangle$, $|1\rangle$ on the Bloch-ball, and the relative phases of the first signal $|S_1\rangle$, and the second signal $|S_2\rangle$:

$$\{|0\rangle, |1\rangle\} = \left\{\frac{1}{\sqrt{2}}(|S_1\rangle + |S_2\rangle), \frac{1}{\sqrt{2}}(|S_1\rangle - |S_2\rangle)\right\}. \tag{18}$$

Hence for example, the $|\nearrow\rangle$ polarization state on the Bloch-ball can be rewritten in the following form:

$$|\nearrow\rangle = \frac{1}{\sqrt{2}}(|0\rangle + |1\rangle) = \frac{1}{\sqrt{2}}(|S_1\rangle + i|S_2\rangle). \tag{19}$$

In the case of $|\nearrow\rangle$, the relative phase between signals $|S_1\rangle$ and $|S_2\rangle$ is $\pi$. As we can conclude, the information encoded in the polarization and in the relative phases are equivalent.

In our geometrical analysis of the DPS QKD protocol, we have used the following conventions of the relative phases of the signals and the polarization states on the Bloch-ball:

$$\begin{aligned}\frac{1}{\sqrt{2}}(|S_1\rangle + |S_2\rangle) &= |0\rangle = |\leftrightarrow\rangle, \\ \frac{1}{\sqrt{2}}(|S_1\rangle + i|S_2\rangle) &= \frac{1}{\sqrt{2}}(|0\rangle + |1\rangle) = |\nearrow\rangle.\end{aligned} \tag{20}$$

In Figure 8 we have illustrated these conventions in the notations of our security analysis. The first and the second signals are denoted by $|S_1\rangle$ and $|S_2\rangle$.

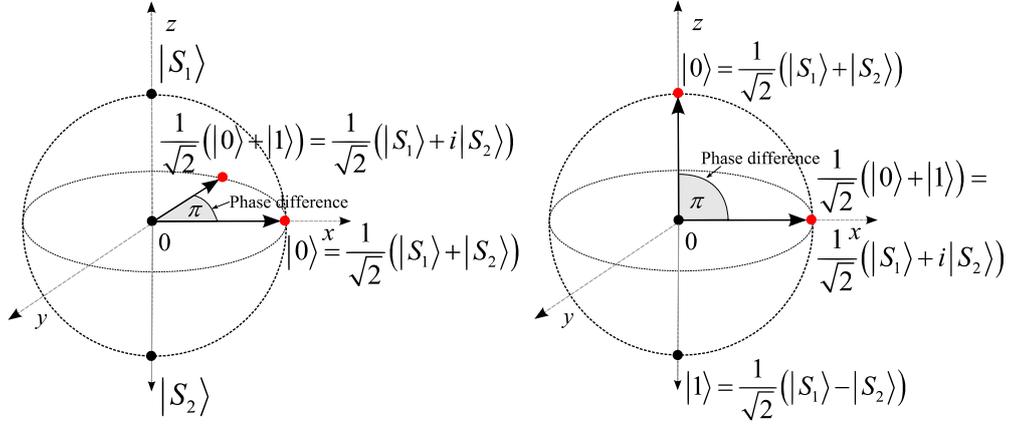

**Figure 8**. The notations used in the geometrical analysis of DPS QKD protocol. The relative phases between the pulses can be represented by polarization angles on the Bloch-ball. The phase encoding and polarization encoding schemes are equivalent mathematically and in our geometrical security analysis.

The soundness of these notations of our geometric analysis is based on the fact that the relative phases between the pulses can be represented by polarization angles on the Bloch-ball, since the phase encoding scheme and the polarization encoding scheme are mathematically the same [1, 2, 3, 19, 20]. Using this equivalence, the time-delay which is caused by the eavesdropper, can be analyzed by the fidelity of her quantum cloner.

Hence, the DPS QKD protocol can be modeled in terms of the polarization states of the B92 protocol. It uses only two polarization states, and the key can be described by a random sequence $B = \left(b_1, b_2, \ldots b_N\right)$ of logical bits, and the generated $N$-length qubit string is:

$$\left|\psi\right\rangle = \left|\psi_{b_1}\right\rangle \otimes \left|\psi_{b_2}\right\rangle \otimes \ldots \otimes \left|\psi_{b_N}\right\rangle = \bigotimes_{i=1}^{N}\left|\psi_{b_i}\right\rangle, \tag{21}$$

where $b_i$ is the basis of the $i$-th qubit [9]. In our geometrical analysis, the $i$-th qubit $\left|\psi_{b_i}\right\rangle$ in the string is generated according to the B92 coding convention, as $\left|\psi_0\right\rangle = \left|\leftrightarrow\right\rangle$ and $\left|\psi_1\right\rangle = \left|\nearrow\right\rangle$. In Figure 9 we illustrated the two polarization states of the DPS QKD protocol, used in our information geometric security analysis.

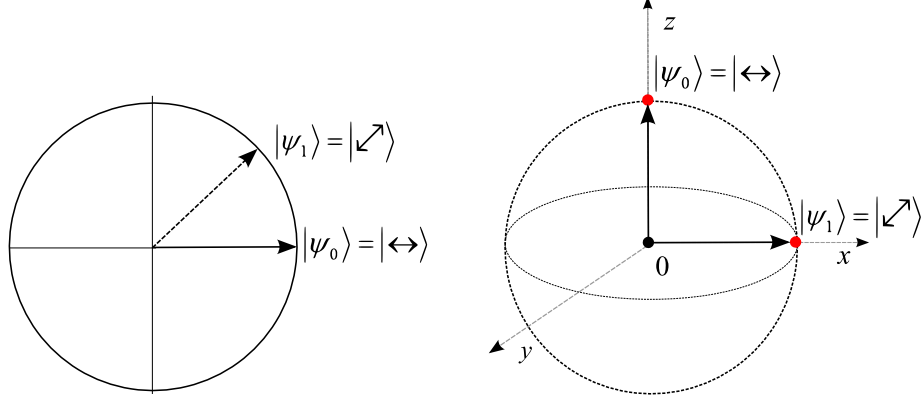

**Figure 9.** The polarization states of B92 protocol.

In general, these polarization states can be expressed by means of some orthogonal basis $\left\{\left|0\right\rangle, \left|1\right\rangle\right\}$ as follows:

$$\left|\pm\alpha\right\rangle = a\left|0\right\rangle \pm b\left|1\right\rangle, \tag{22}$$

where the coefficients can be given by

$$a = \sqrt{\frac{1}{2}\left(1 + e^{-2\mu_\alpha}\right)} \text{ and } b = \sqrt{\frac{1}{2}\left(1 - e^{-2\mu_\alpha}\right)}, \tag{23}$$

and $a, b \in \mathbb{R}$ and $a^2 + b^2 = 1$, and $a > b$ if $\mu_\alpha \neq 0$.

In Figure 10 we show the polarization states of $\left|\pm\alpha\right\rangle = a\left|0\right\rangle \pm b\left|1\right\rangle$ in the Bloch-ball representation, and we depict the $\pi$ phase difference between quantum states $\left\{\left|\rho_1\right\rangle, \left|\rho_2\right\rangle\right\}$ and $\left\{\left|\rho_3\right\rangle, \left|\rho_4\right\rangle\right\}$. In practice, Alice sends a WCP signal, whose phases

are randomly modulated by 0 or $\pi$. In our security analysis, these WCP signals are modeled by the quantum states $\{|\rho_1\rangle,|\rho_2\rangle\}$ and $\{|\rho_3\rangle,|\rho_4\rangle\}$ on the Bloch-ball.

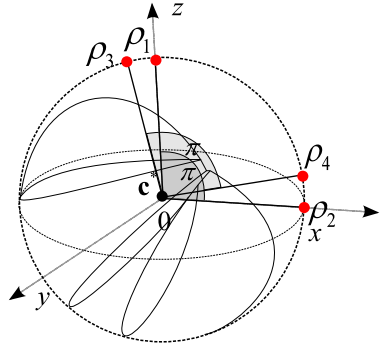

**Figure 10.** The WCP pulses of DPS QKD protocol are modeled by different polarization states.

In Figure 11 we show the result of the eavesdropper's attack. For the best results, Eve uses the universal cloner for non-equatorial states $\rho_1, \rho_2$ and $\rho_3$, and the phase-covariant cloner for equatorial states. The Delaunay tessellation of these cloned states is different from the Delaunay tessellation of pure states, and the radius of the smallest enclosing quantum informational ball will be also different [7, 16].

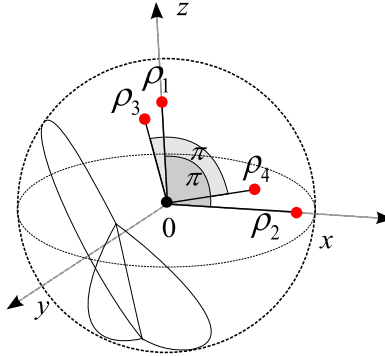

**Figure 11.** The tessellation of the Bloch-ball for cloned quantum states differs from the diagram of the pure states originally sent.

As can be concluded, the smallest enclosing quantum informational ball contains all the cloned states. The length of the radius of the smallest quantum informational ball describes the eavesdropper's maximally obtainable information. In Figure 12 we show the smallest enclosing quantum informational ball and the convex hull of the quantum states.

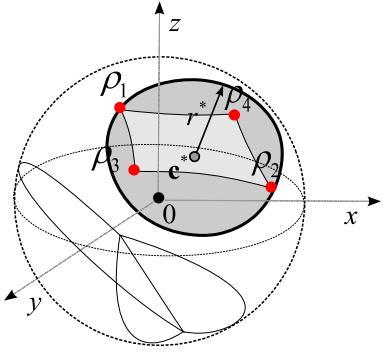

**Figure 12.** The radius of the smallest quantum informational ball describes the eavesdropper's maximum obtainable information. The algorithm computes the length of the information-theoretical radius by the determination of the convex hull of mixed quantum states.

In the next section we show the methods for convex hull calculation, and our techniques for the determination of smallest quantum informational ball.

### 3.4 Computation of Delaunay Triangulation on Bloch sphere

We would like to compute the information-theoretical radius $r^*$ of the smallest enclosing ball of the cloned mixed quantum states which describes the eavesdropper's maximal obtainable information, thus we must first seek the center $\mathbf{c}^*$ of the set of quantum states $\mathcal{S}$. The set $\mathcal{S}$ of quantum states is denoted by $\mathcal{S} = \{\rho_i\}_{i=1}^n$. The distance $d(\cdot,\cdot)$ between any two quantum states of $\mathcal{S}$ is measured by the quantum relative entropy, thus a minimax mathematical optimization is applied to the quantum relative entropy-based distances to find the center $\mathbf{c}$ of the set $\mathcal{S}$. We denote the quantum relative entropy from $\mathbf{c}$ to the furthest point of $\mathcal{S}$ by $d(\mathbf{c},\mathcal{S}) = \max_i d(\mathbf{c},\rho_i)$. Using a minimax optimization, we can minimize the maximal quantum relative entropy from $\mathbf{c}$ to the furthest point of $\mathcal{S}$ by $\mathbf{c}^* = \arg\min_{\mathbf{c}} d(\mathbf{c},\mathcal{S})$. In Figure 13, we have illustrated the circumcenter $\mathbf{c}^*$ of $\mathcal{S}$ for the Euclidean distance and for quantum relative entropy [5, 17].

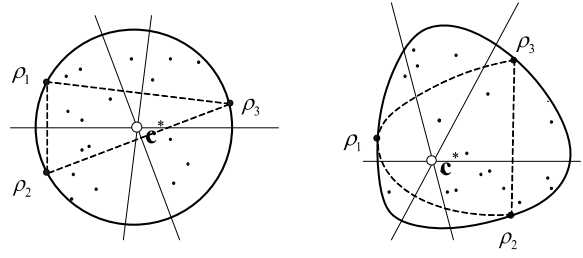

**Figure 13.** Circumcenter for Euclidean distance and quantum relative entropy.

In classical computational geometry, Voronoi diagrams and Delaunay triangulations play an important role [7, 16]. A Voronoi diagram is a division of space. The dual diagram for a Voronoi diagram is called a Delaunay tessellation [7, 16]. In the graph of a Delaunay triangulation, any circle is empty if it contains no vertex of $\mathcal{S}$ in its interior. If two quantum states of set $\mathcal{S}$ are denoted by $\rho$ and $\sigma$, then edge $e$ is in $Del(\mathcal{S})$ if and only if there exists an empty circle that passes through $\rho$ and $\sigma$. An edge satisfying the empty circle property is said to be Delaunay. The Delaunay triangulation is guaranteed to be a triangulation only if the vertices of $\mathcal{S}$ are in a general position, thus there are no four quantum states of $\mathcal{S}$ lying on the same circle. The circumcircle of a triangle is the unique circle that passes through all three of its vertices, and the triangle is Delaunay if and only if its circumcircle is empty. The quantum Delaunay triangulation of a set of quantum states $\mathcal{S}$, denoted by $Del(\mathcal{S})$, is the geometric dual of quantum Voronoi diagrams $vo(\mathcal{S})$. In Figure 14 we compare a classical Euclidean Delaunay and a quantum informational Delaunay triangulation for a set of quantum states.

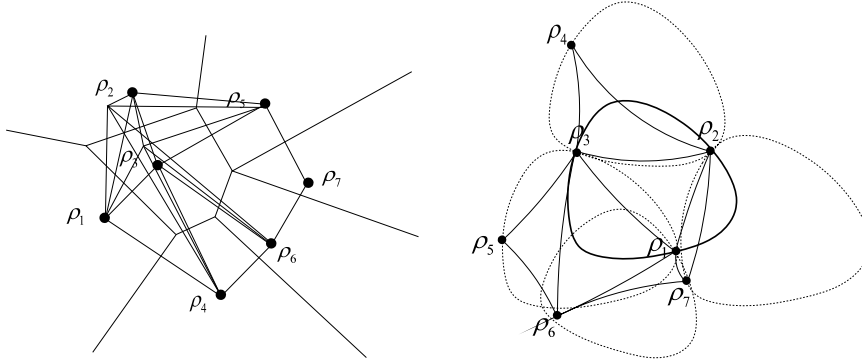

**Figure 14.** Comparison of classical Euclidean (a) and quantum Delaunay triangulation (b).

The quantum Voronoi diagrams can be first-type or right-sided diagrams. Similarly, we can derive two triangulations from quantum Voronoi diagrams. The first-type quantum informational ball circumscribing any simplex of quantum Delaunay triangulation $Del(\mathcal{S})$ is empty. If we choose a subset $\Upsilon$ of at most $d+1$ states in

$\mathcal{S} = \{\rho_1,...,\rho_n\}$, then the convex hull of the associated quantum states $\rho_i, i \in \Upsilon$, is a simplex of the quantum triangulation of $\mathcal{S}$, iff there exists an empty quantum informational ball $B$ passing through the $\rho_i$, $i \in \Upsilon$. The first-type and second-type quantum diagrams for quantum states which have non-equal radii, are different. In Figure 15, we compare the first-type and second-type quantum Delaunay diagrams for mixed quantum states on the Bloch sphere.

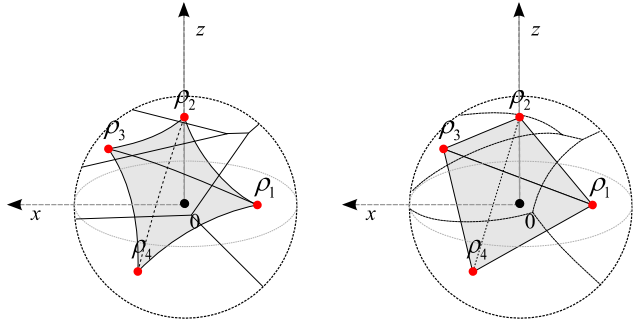

**Figure 15.** Comparison of first-type and second-type quantum Delaunay triangulations on the Bloch-ball.

The security of the quantum channel is determined by our geometrical model based on the assumptions $r^* > r^*_{Eve}$, and $r^* \leq r^*_{Eve}$, and the approximate value of the fidelity parameter $F_{Eve}$, can be expressed as:

$$F_{Eve} = \langle\psi|^{(in)} \rho^{(out)} |\psi\rangle^{(in)} = \frac{1}{2}(1+r). \tag{24}$$

The approximated value of the information-theoretical impacts of the eavesdropper is obtained by $r^*$, the radius of the smallest information ball.

## 4 Geometrical Representation of Quantum States

The quantum relative entropy for general quantum state $\rho = (x,y,z)$ and mixed state $\sigma = (\tilde{x},\tilde{y},\tilde{z})$, with radii $r_\rho = \sqrt{x^2+y^2+z^2}$ and $r_\sigma = \sqrt{x^2+y^2+z^2}$ is given by

$$D(\rho\|\sigma) = \frac{1}{2}\log\frac{1}{4}(1-r_\rho^2) + \frac{1}{2}r_\rho \log\frac{(1+r_\rho)}{(1-r_\rho)} - \frac{1}{2}\log\frac{1}{4}(1-r_\sigma^2) - \frac{1}{2r_\sigma}\log\frac{(1+r_\sigma)}{(1-r_\sigma)}\langle\rho,\sigma\rangle, \tag{25}$$

where $\langle \rho, \sigma \rangle = (x\tilde{x} + y\tilde{y} + z\tilde{z})$. For a maximally mixed state $\sigma = (\tilde{x}, \tilde{y}, \tilde{z}) = (0,0,0)$ and $r_\sigma = 0$, the quantum relative entropy can be expressed as

$$D(\rho\|\sigma) = \frac{1}{2}\log\frac{1}{4}(1 - r_\rho^2) + \frac{1}{2}r_\rho \log\frac{(1 + r_\rho)}{(1 - r_\rho)} - \frac{1}{2}\log\frac{1}{4}. \quad (26)$$

The relative entropy of quantum states can be described by a strictly convex and differentiable generator function $\mathbf{F}$:

$$\mathbf{F}(\rho) = -\mathsf{S}(\rho) = Tr(\rho \log \rho), \quad (27)$$

where $-\mathsf{S}$ is the negative entropy of quantum states. The quantum relative entropy $D(\rho\|\sigma)$ for density matrices $\rho$ and $\sigma$ is given by generator function $\mathbf{F}$ in the following way:

$$D(\rho\|\sigma) = \mathbf{F}(\rho) - \mathbf{F}(\sigma) - \langle \rho - \sigma, \nabla \mathbf{F}(\sigma) \rangle, \quad (28)$$

where $\langle \rho, \sigma \rangle = Tr(\rho\sigma^*)$ is the inner product of quantum states and $\nabla \mathbf{F}(\cdot)$ is the gradient.

In Figure 16, we have depicted the quantum informational distance, $D(\rho\|\sigma)$, as the vertical distance between the generator function $\mathbf{F}$ and $H(\sigma)$, the hyperplane tangent to $\mathbf{F}$ at $\sigma$. The point of intersection of quantum state $\rho$ on $H(\sigma)$ is denoted by $H_\sigma(\rho)$.

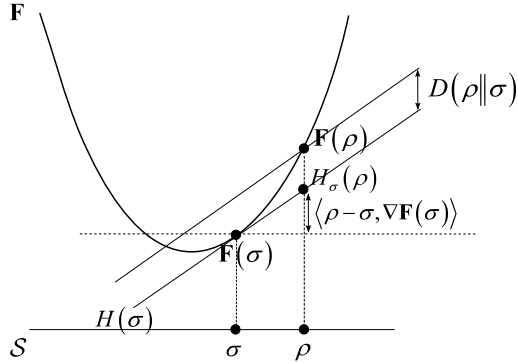

**Figure 16.** Depiction of generator function as a negative von Neumann entropy.

For the quantum informational distance function, the generator function is the negative von Neumann entropy function $-\mathsf{S}$,

$$\mathbf{F}(\rho) = -\mathsf{S}(\rho) = Tr(\rho \log \rho), \quad (29)$$

where $F : S(\mathbb{C}^d) \to \mathbb{R}$. The quantum informational distance function $D_\mathbf{F}(\rho\|\sigma)$ with generator function $\mathbf{F}(\rho) = -\mathsf{S}(\rho)$ is illustrated in Figure 17 [31-36].

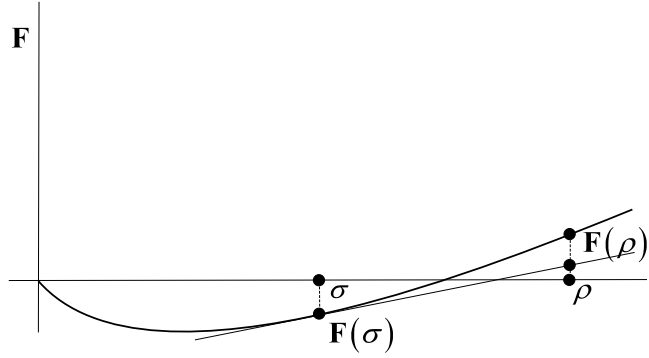

**Figure 17.** Negative von Neumann generator function.

The quantum informational distance has some distance-like properties, however it is not commutative [4, 6, 7], thus $D(\rho\|\sigma) \neq D(\sigma\|\rho)$, and $D(\rho\|\sigma) \geq 0$ iff $\rho \neq \sigma$, and $D(\rho\|\sigma) = 0$ iff $\rho = \sigma$.

### 4.1 Laguerre Diagram for Quantum States

We use the Laguerre Delaunay diagram [7, 8] to compute the radius of the smallest enclosing ball. In general, the Laguerre distance for generating points $x_i$ with weight $r_i^2$, in a Euclidean space is defined by [33-35]

$$d_L(\rho, x_i) = \|\rho - x_i\|^2 - r_i^2. \quad (30)$$

The Delaunay diagram for the Laguerre distance is called the Laguerre-Delaunay diagram. For the Laguerre bisector of two three-dimensional Euclidean balls $B(\rho, r_P)$ and $B(\sigma, r_Q)$ centered at quantum states $\rho$ and $\sigma$, we can write the equation [31-36]

$$2\langle x, \sigma - \rho \rangle + \langle \rho, \rho \rangle - \langle \sigma, \sigma \rangle + r_Q^2 - r_P^2 = 0. \quad (31)$$

In a Euclidean space, the Laguerre distance $d_L(\rho, x_i)$ with weight $r_i^2$ can be interpreted as the square of the length of the line segment starting at $\rho$ and tangent to the circle centered at $x_i$ with radius $\sqrt{r_i^2}$. Thus, the circle centered at $x_i$ with radius $\sqrt{r_i^2}$ is the circle associated with $x_i$ [7, 8], [31-36].

We show a new method for deriving the quantum relative entropy-based Delaunay tessellation on the Bloch-ball $\mathcal{B}$ to detect eavesdropping activity on the quantum channel. In our algorithm we present an effective solution to seek the center **c** of the set of smallest enclosing quantum information ball, using *Laguerre* diagrams [7, 8], [31-36].

Our geometrical-based security analysis has two main steps:

1. We construct Delaunay triangulation from Laguerre diagrams on the Bloch-ball.
2. We seek the center of the smallest enclosing ball.

### 4.2 Quantum Delaunay Triangulation from Laguerre Diagrams

As we have seen, in a Euclidean space, the Laguerre distance of a point $x$ to a Euclidean ball $b = b(\rho, r)$ is defined as $d_L(\rho, x) = \|\rho - x\|^2 - r^2$, and for $n$ balls $b_i = b(\rho_i, r_i)$, where $i = 1, \ldots, n$, the Laguerre diagram [7, 8] of $b_i$ is defined as the minimization diagram of the corresponding $n$ distance functions

$$d_L^i(x) = \|\rho - x\|^2 - r^2. \tag{32}$$

In Figure 18, we show the ordinary triangulation of quantum relative entropy-based Voronoi diagram [31-33].

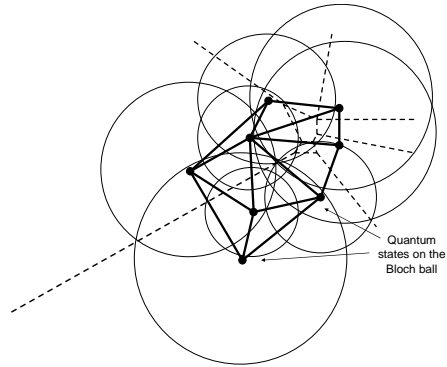

**Figure 18.** Regular triangulation on the Bloch-ball.

We use the result of Aurenhammer to construct the quantum relative entropy-based dual diagram of the Delaunay tessellation, using the Laguerre diagram of the $n$ Euclidean spheres of equations [7]

$$\langle x - \rho_i', x - \rho_i' \rangle = \langle \rho_i', \rho_i' \rangle + 2\left(\mathbf{F}(\rho_i') - \langle \rho_i, \rho_i' \rangle\right). \tag{33}$$

The most important result of this equivalence is that we can efficiently construct a quantum relative entropy-based Delaunay triangulation on the Bloch sphere, using fast methods for constructing classical Euclidean Laguerre diagrams [7, 8].

### 4.3 Center of the Quantum Informational Ball

In our security analysis we use an approximation algorithm from classical computational geometry to determine the smallest enclosing ball of balls using core-sets. The core-sets have an important role in our calculation and approximate method. We apply the approximation algorithm presented by Badoui and Clarkson, however in our algorithm the distances between quantum states are measured by quantum relative entropy [13]. The $\mathcal{E}$-core set $\mathcal{C}$ is a subset of the set $\mathcal{C} \subseteq \mathcal{S}$, such that for the circumcenter $\mathbf{c}$ of the minimax ball [5]

$$d(\mathbf{c}, \mathcal{S}) \leq (1 + \mathcal{E})r, \qquad (34)$$

where $r$ is the radius of the smallest enclosing quantum information ball of the set of quantum states $\mathcal{S}$ [13]. The approximating algorithm, for a set of quantum states $S = \{s_1, \ldots, s_n\}$ and circumcenter $\mathbf{c}$, first finds the farthest point $s_m$ of ball set $B$, and moves $\mathbf{c}$ towards $s_m$ in $\mathcal{O}(dn)$ time in every iteration step.

The algorithm seeks the farthest point in the ball set $B = \{b_1 = Ball(\mathbf{c}_1, r_1), \ldots, b_n = Ball(\mathbf{c}_n, r_n)\}$ by maximizing the quantum informational distance for a current circumcenter position $\mathbf{c}$ as $\max_{i \in \{1,\ldots,n\}} D_F(\mathbf{c}, b_i)$. Using equation $\max_{x \in b_i} D_F(\mathbf{c}, x_i) = D_F(\mathbf{c}, S_i) + r_i$, we get

$$\max_{i \in \{1,\ldots,n\}} D_F(\mathbf{c}, b_i) = \max_{i \in \{1,\ldots,n\}} (D_F(\mathbf{c}, S_i) + r_i). \qquad (35)$$

In Figure 19, we illustrate the smallest enclosing ball of balls in the quantum space [32-35].

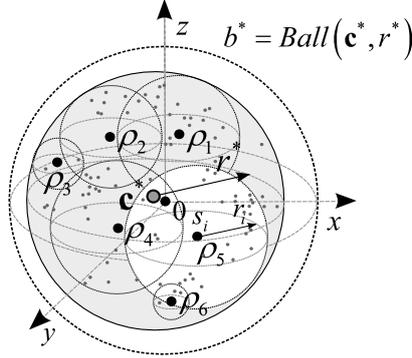

**Figure 19.** The smallest enclosing ball of a set of balls in the quantum space.

We denote the set of $n$ $d$-dimensional balls by $B = \{b_1, \ldots, b_n\}$, where $b_i = Ball(S_i, r_i)$, $S_i$ is the center of ball $b_i$ and $r_i$ is the radius of the $i$-th ball. The smallest enclosing ball of set $B = \{b_1, \ldots, b_n\}$ is the unique ball $b^* = Ball(\mathbf{c}^*, r^*)$ with minimum radius $r^*$ and center $\mathbf{c}^*$ [13]. The algorithm does $\left\lfloor \dfrac{1}{\mathcal{E}^2} \right\rfloor$ iterations to ensure an $(1 + \mathcal{E})$ approximation, thus the overall cost of the algorithm is $\mathcal{O}\left(\dfrac{dn}{\mathcal{E}^2}\right)$ [5]. The smallest enclosing ball of ball set $B$ can be written as

$$\min_{\mathbf{c}} \mathbf{F}_B(\mathbf{c}), \tag{36}$$

where $\mathbf{F}_B(X) = d(X,B) = \max_{i \in \{1,\ldots,n\}} d(X,B_i)$ and the distance function $d(\cdot,\cdot)$ measures the relative entropy between quantum states [6, 13]. The minimum ball of the set of balls [33-36] is unique, thus the circumcenter $\mathbf{c}^*$ of the set of quantum states is $\mathbf{c}^* = \arg\min_{\mathbf{c}} \mathbf{F}_B(\mathbf{c})$.

The main steps of our algorithm can be summarized as [34-36]:

**Algorithm 1.**
1. *Select* a random center $\mathbf{c}_1$ from the set of quantum states $\mathcal{S}$
$$\mathbf{c}_1 = S_1$$
**for** $\left(i = 1, 2, \ldots, \left\lceil \frac{1}{\mathcal{E}^2} \right\rceil \right)$
**do**
2. *Find* the farthest point $s$ of $\mathcal{S}$ wrt. quantum relative entropy
$$S \leftarrow \arg\max_{s' \in \mathcal{S}} D_F(\mathbf{c}_i, s')$$
3. *Update* the circumcircle:
$$\mathbf{c}_{i+1} \leftarrow \nabla_F^{-1}\left(\frac{i}{i+1}\nabla_F(\mathbf{c}_i) + \frac{1}{i+1}\nabla_F(S)\right).$$
4. *Return* $\mathbf{c}_{i+1}$

At the end of our algorithm, the radius $r^*$ of the smallest enclosing ball with respect to the quantum informational distance is equal to the information-theoretical fidelity of the cloning transformation.

# 5 Fitting the Smallest Quantum Informational Ball

Geometrically, the smallest quantum informational ball can be computed from the intersection of contours of the quantum relative entropy ball with the ellipsoid of the secret channel, which ellipsoid is generated by the eavesdropper's cloner machine. The maximum length radius $\mathbf{r}_\rho$ can be determined by an iterative algorithm, using the quantum relative entropy as a distance measure [41-42].

In Figure 20(a), the smallest quantum informational ball with radius $r^* = D_{\max}(\mathbf{r}_\rho \| \mathbf{r}_\sigma)$ intersects the channel ellipsoid at magnitude $m_\rho$ of the Bloch vector $\mathbf{r}_\rho$. The Euclidean distance between the origin and center $\mathbf{c}^*$ is denoted by $m_\sigma$. Similarly, the Euclidean distance between the origin and state $\rho$ is denoted by $m_\rho$. In our geometrical iteration algorithm, we would like to determine the location of vector $\mathbf{r}_\sigma$ inside the channel ellipsoid such that, the largest possible contour value $D_{\max}(\mathbf{r}_\rho \| \mathbf{r}_\sigma)$ touches the channel ellipsoid surface and the remainder of the $D_{\max}$

contour surface lies entirely outside the channel ellipsoid. The point on the channel ellipsoid surface is defined as the set of channel output $\rho$. The vector $\mathbf{r}_\sigma$ is defined in the interior of the ellipsoid, as the convex hull of the channel ellipsoid. To determine the optimal length of the radius, the algorithm moves point $\sigma$.

As we move vector $\mathbf{r}_\sigma$ from the optimum position, a larger contour corresponding to the larger value of the quantum relative entropy $D$ will intersect the channel ellipsoid surface, thereby $\max_{\mathbf{r}_\rho} D(\mathbf{r}_\rho \| \mathbf{r}_\sigma)$ will increase. We can conclude that vector $\mathbf{r}_\sigma$ should be adjusted to minimize $\max_{\mathbf{r}_\rho} D(\mathbf{r}_\rho \| \mathbf{r}_\sigma)$, as illustrated in Figure 20(b).

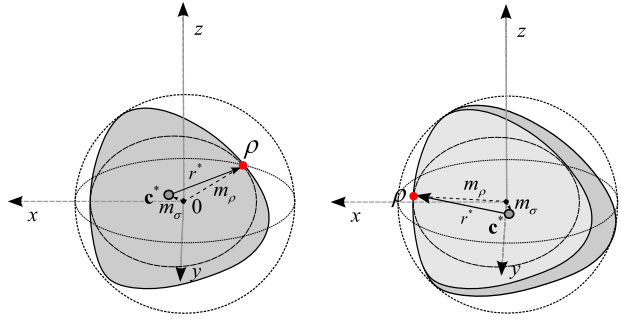

**Figure 20.** Intersection of radius of smallest enclosing quantum informational ball and channel ellipsoid (a). The optimal ball is shown in light-grey (b).

The computed radius is equal to the radius of the smallest quantum informational ball, hence the quantum informational radius can be used to derive the fidelity of the eavesdropper's quantum cloner machine. The vector $\mathbf{r}_\sigma$ should be adjusted to minimize the value of $\max_{\mathbf{r}_\rho} D(\mathbf{r}_\rho \| \mathbf{r}_\sigma)$. To find the optimal value of vector $\mathbf{r}_\sigma$ in our geometrical analysis, we choose a start point for vector $\mathbf{r}_\sigma$ in the interior of the ellipsoid.

In Figure 21(a), we show the initial start point inside the channel ellipsoid chosen by the algorithm. The vector of state $\sigma$ is denoted by $\mathbf{r}_\sigma$. In the next step, the algorithm determines the set of points to the vector $\mathbf{r}'_\rho$ on the ellipsoid surface most distant from $\mathbf{r}_\sigma$, using the quantum relative entropy as distance measure. In Figure 21(b), the new state is notated by $\rho'$.

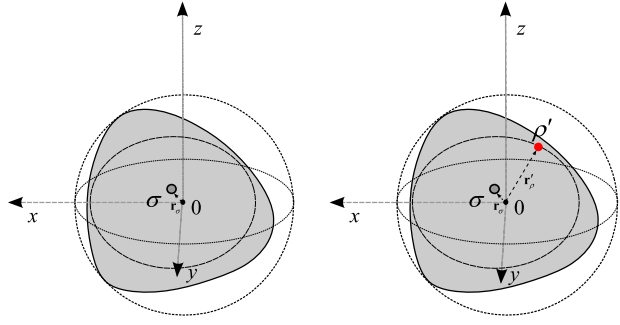

**Figure 21.** The algorithm determines the points on the ellipsoid surface most distant from the point, using the quantum relative entropy as distance measure.

The maximum distance between the states can be expressed as $\max_{\mathbf{r}_\rho} D\left(\mathbf{r}'_\rho \| \mathbf{r}_\sigma\right)$. We choose a random Bloch sphere vector from the maximal set of points according to vector $\mathbf{r}'_\rho$. The selected point is denoted by $\mathbf{r}''_\rho$. The algorithm makes a step from $\mathbf{r}_\sigma$ towards the surface point vector $\mathbf{r}''_\rho$ in the Bloch sphere space. In this step, the algorithm updates vector $\mathbf{r}_\sigma$ to

$$\mathbf{r}^*_\sigma = (1-\gamma)\mathbf{r}_\sigma + \gamma \mathbf{r}''_\rho, \tag{37}$$

where $\gamma$ denotes the size of the step. In Figure 22(a), the updated state and the vector of the state are denoted by $\rho''$ and $\mathbf{r}''_\rho$. The center of the quantum informational ball is denoted by $\mathbf{r}^*_\sigma$. In Figure 22(b), we illustrate the quantum informational distance between the final center point and the maximal distance state $\rho$, using the notation $\max_{\mathbf{r}_\rho} D\left(\mathbf{r}_\rho \| \mathbf{r}_\sigma\right)$. Using the updated vector $\mathbf{r}^*_\sigma$, the algorithm continues to loop until $\max_{\mathbf{r}^*_\rho} D\left(\mathbf{r}'_\rho \| \mathbf{r}^*_\sigma\right)$ no longer changes. We conclude that the iteration converges to the optimal $\mathbf{r}_\sigma$, because the algorithm minimizes $\max_{\mathbf{r}_\rho} D\left(\mathbf{r}_\rho \| \mathbf{r}_\sigma\right)$.

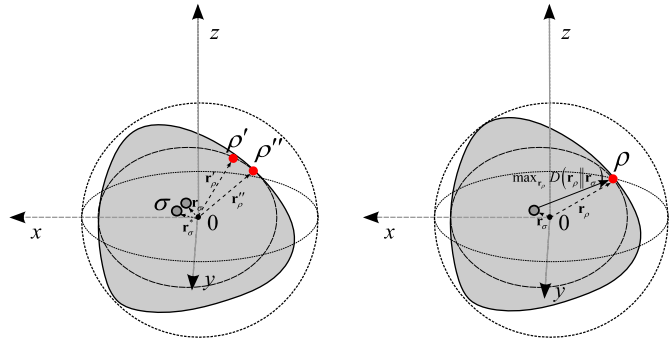

**Figure 22.** The algorithm makes a step towards the found surface point vector and updates the vector.

At the end of the iteration process, the radius of the smallest quantum informational ball can be expressed as

$$\min \max_{\mathbf{r}_\rho} D(\mathbf{r}_\rho \| \mathbf{r}_\sigma). \tag{38}$$

In Figure 23, we compare the smallest quantum informational ball and the ordinary Euclidean ball.

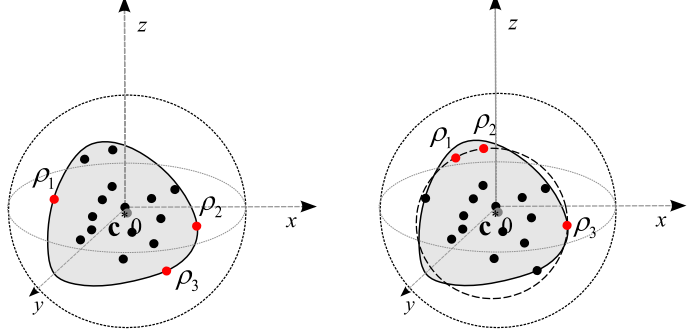

**Figure 23.** The maximum distance states of the smallest balls differ for the quantum informational distance and Euclidean distance.

We can conclude that the quantum states $\rho_1, \rho_2$ and $\rho_3$, which determine the Euclidean smallest enclosing ball are different from the states of the quantum informational ball.

### 5.1 Smallest Quantum Informational Ball for UCM-Based Cloning

The UCM cloner-based attack in the DPS QKD protocol can be detected if the radius $r_\mathcal{E}$ of imperfect UCM cloning is equal or greater than the radius $r_{\mathcal{E},UCM}$ of the idealistic UCM ball in ellipsoid $\mathcal{E}$ representation, thus

$$r_\mathcal{E} \geq r_{\mathcal{E},UCM} = \sqrt{x_\mathcal{E}^2 + y_\mathcal{E}^2 + z_\mathcal{E}^2} = \sqrt{3\frac{1}{12}} = \frac{1}{2}. \tag{39}$$

This surface is an oblate ellipsoid $\mathcal{E}$ and can be expressed by $\mathcal{E} = x^2 + y^2 + z^2 + xy + xz + yz = \frac{1}{2}$. The ellipsoid $\mathcal{E}(x_\mathcal{E}, y_\mathcal{E}, z_\mathcal{E})$ has polar radius $x_\mathcal{E} = \frac{1}{2}$, while the equatorial radius is $z_\mathcal{E} = 1$. The distance to the origin is $x_\mathcal{E}^2 + y_\mathcal{E}^2 + z_\mathcal{E}^2 = p_x + p_y + p_z$, thus the closest point to the origin is at the pole of the ellipsoid $\mathcal{E}$ and can be expressed as [9, 10, 11, 12]

$$\left(\frac{1}{\sqrt{12}}, \frac{1}{\sqrt{12}}, \frac{1}{\sqrt{12}}\right). \tag{40}$$

Using the ellipsoid $\mathcal{E}$ representation, we can model the effects of Eve's quantum cloner. The cloning transformation will be detected by Bob, if the point $(x_\mathcal{E}, y_\mathcal{E}, z_\mathcal{E})$ representing the quality of the cloning transformation, lies on or outside the optimal UCM ball, represented by ellipsoidal radius $r_{\mathcal{E},UCM}$.

In Figure 24, we illustrate the radius $r_{\mathcal{E},UCM}$ of the UCM ball and the radius $r_\mathcal{E}$ of the corresponding imperfect UCM cloning transformation. The origin of $\mathcal{E}$ represents zero cloning activity in the channel.

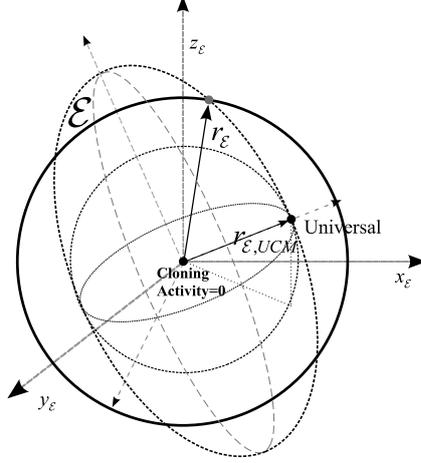

**Figure 24.** Comparison of optimal UCM and imperfect universal cloning.

In our quantum informational distance-based geometrical security analysis, Bob will detect the quantum cloner, if $r_\mathcal{E} \geq r_{\mathcal{E},UCM}$, because in this case we can give the following condition for the radius $r^*$ of his smallest enclosing quantum informational ball:

$$r^* \leq 1 - \mathsf{S}\left(1 - \frac{4(r_\mathcal{E})^2}{3}\right) \leq 1 - \mathsf{S}\left(1 - \frac{4(r_{\mathcal{E},UCM})^2}{3}\right). \tag{41}$$

In this geometrical representation - if there is no quantum cloner on the quantum channel - then $r_\mathcal{E} = 0$, thus in this case Bob has a quantum informational ball with radius $r^* = 1$.

In Figure 25, we show the information-theoretical radii $r^*$ and $r^*_{UCM}$. The smallest enclosing quantum ball of the imperfect UCM cloner has radius

$$r^* = 1 - \mathsf{S}\left(1 - \frac{4(r_\mathcal{E})^2}{3}\right), \tag{42}$$

while the radius of the idealistic UCM-based cloning attack in the DPS QKD protocol can be expressed as

$$r^*_{UCM} = 1 - \mathsf{S}\left(1 - \frac{4\left(r_{\mathcal{E},UCM}\right)^2}{3}\right). \tag{43}$$

The smallest quantum informational ball with radius $r^*$ is shown in grey, the ball of the idealistic UCM cloner with radius $r^*_{UCM}$ is shown in light grey.

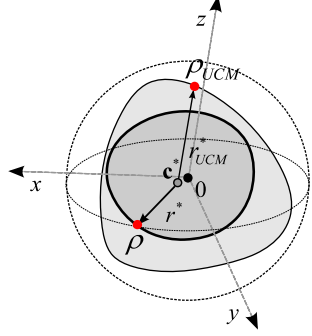

**Figure 25.** The smallest enclosing quantum informational ball of optimal and imperfect universal cloner.

We can conclude that, if $r_{\mathcal{E}} \geq r_{\mathcal{E},UCM}$, then $r^* \leq r^*_{UCM}$, hence the information-theoretical radius will be smaller.

## 5.2 Smallest Quantum Informational Ball for Phase-covariant Based Attack

In the phase-covariant based attack against the QKD quantum cryptography protocol, the cloning activity can be detected by Bob, if the radius $r_{\mathcal{E}}$ of the *imperfect* phase-covariant cloner is equal or greater than the radius $r_{\mathcal{E},phasecov}$ of the phase-covariant ball in the ellipsoid $\mathcal{E}$ representation, $r_{\mathcal{E}} \geq r_{\mathcal{E},phasecov}$.

Using the ellipsoid $\mathcal{E}$ representation, we can model the effects of Eve's phase-covariant quantum cloner-based attack. The imperfect cloning transformation is denoted by point $\left(x_{\mathcal{E}}, 0, z_{\mathcal{E}}\right)$, which lies on or outside the optimal phase-covariant ball.

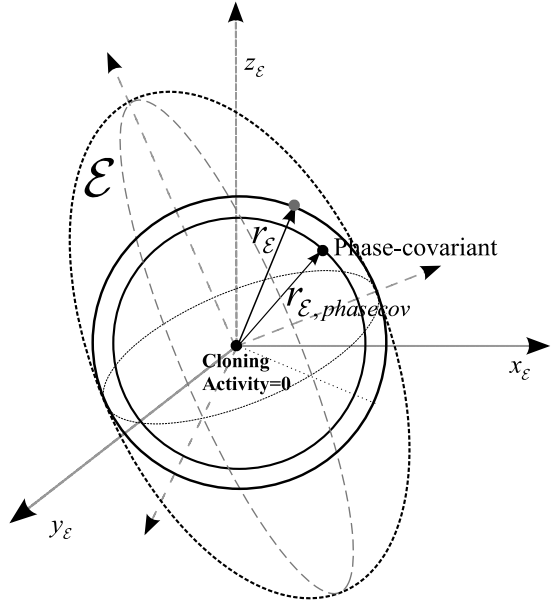

**Figure 26.** The ellipsoidal radii for optimal phase-covariant cloning and imperfect cloning activity.

The local coordinate system $\left(x_{\mathcal{E}},0,z_{\mathcal{E}}\right)$ represents the quality of the cloning transformation, and the eavesdropping activity will be detected by Bob, if

$$r_{\mathcal{E}} = \sqrt{x_{\mathcal{E}}^{2} + 0 + z_{\mathcal{E}}^{2}} \geq \left(\frac{2}{3} - \frac{4}{3\sqrt{8}}\right). \tag{44}$$

Bob will detect the quantum cloner if $r_{\mathcal{E}} \geq r_{\mathcal{E},phasecov}$, where $r_{\mathcal{E}}$ is the radius representing the imperfect phase-covariant cloning attack. In this case, we can give the following condition for the information-theoretical radius $r^{*}$ of his smallest quantum informational ball

$$r^{*} \leq 1 - \mathsf{S}\left(1 - \frac{3\left(r_{\mathcal{E}}\right)^{2}}{2}\right) \leq 1 - \mathsf{S}\left(1 - \frac{3\left(r_{\mathcal{E},phasecov}\right)^{2}}{2}\right) = 1 - \mathsf{S}\left(\frac{1}{2} + \sqrt{\frac{1}{8}}\right). \tag{45}$$

The smallest quantum informational ball with radius $r^{*}$ is shown in grey, the maximal ball of the phase-covariant cloner is shown in light grey. In the figures the information-theoretical radius is denoted by $r^{*}$. The quantum ball of the imperfect phase-covariant cloner is illustrated with radius

$$r^{*} = 1 - \mathsf{S}\left(1 - \frac{3\left(r_{\mathcal{E}}\right)^{2}}{2}\right), \tag{46}$$

the idealistic phase-covariant cloner is denoted by radius

$$r^*_{phasecov} = 1 - \mathsf{S}\left(1 - \frac{3\left(r_{\mathcal{E},phasecov}\right)^2}{2}\right). \tag{47}$$

In Figure 27, we compare an idealistic phase-covariant cloner quantum ball and an imperfect phase-covariant cloner quantum ball.

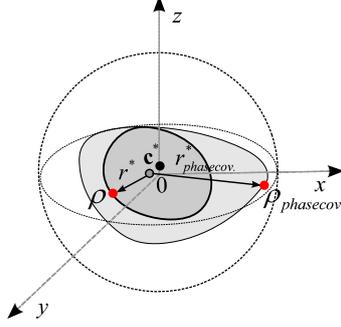

**Figure 27.** The smallest enclosing quantum informational ball of optimal and imperfect phase-covariant cloner.

It can be concluded that the information-theoretical radii for idealistic and imperfect phase-covariant cloning are different.

### 5.3 Comparison of UCM and Phase-covariant Based Attacks

In the three-dimensional $\mathcal{E}$ ellipsoid representation, the radius $r_{\mathcal{E},phasecov}$ of the phase-covariant cloner-based attack is smaller than radius $r_{\mathcal{E},UCM}$. For the radius of the UCM and phase-covariant ball in the ellipsoidal $\mathcal{E}$ representation

$$r_{\mathcal{E},phasecov} < r_{\mathcal{E},UCM}, \tag{48}$$

where $r_{\mathcal{E},phasecov} = \sqrt{x_{\mathcal{E}}^2 + y_{\mathcal{E}}^2 + z_{\mathcal{E}}^2} = \sqrt{\frac{2}{3} - \frac{4}{3\sqrt{8}}}$.

In Figure 28, we illustrate $r_{\mathcal{E},phasecov}$ and $r_{\mathcal{E},UCM}$ in the three-dimensional ellipsoidal representation.

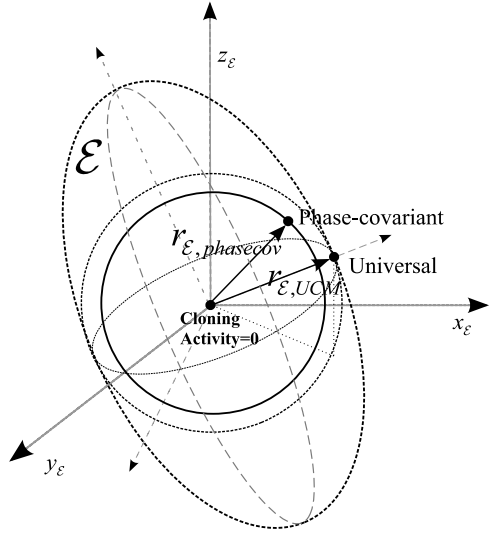

**Figure 28.** Comparison of UCM and phase-covariant based attack in ellipsoidal representation.

Using the results derived in Section 3.1, the following connection holds between radii $r^*_{UCM}$ and $r^*_{phasecov}$ of the smallest enclosing quantum informational balls of UCM and phase-covariant cloning-based attack:

$$r^*_{UCM} = 1 - \mathsf{S}\left(1 - \frac{4\left(r_{\mathcal{E},UCM}\right)^2}{3}\right) \leq r^*_{phasecov} = 1 - \mathsf{S}\left(1 - \frac{3\left(r_{\mathcal{E},phasecov}\right)^2}{2}\right). \tag{49}$$

In Figure 29, we show the radii $r^*_{UCM}$ and $r^*_{phasecov}$ of the smallest enclosing quantum informational ball for a UCM-based attack and for DPS QKD protocol, in the Bloch sphere representation.

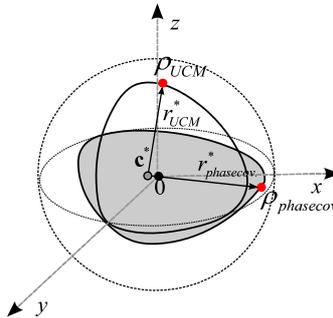

**Figure 29.** Comparison of smallest enclosing quantum informational ball of UCM and phase-covariant cloners.

It can be concluded that the best quality of the two outputs simultaneously can be realized with a UCM. If an eavesdropper uses a phase-covariant cloner, one of the two outputs should have better fidelity, while the fidelity of the second output will be lower.

### 5.4 Description of Different Cloner Machines for the DPS QKD

Using the results derived in Sections 5.1 and 5.2, the quantum channel in the DPS QKD protocol is secure if conditions $r^* > r^*_{phasecov}$ and $r^* > r^*_{UCM}$ hold for the radius of the smallest enclosing quantum informational ball. In our geometrical method, we compute $r^*$, the radius of the smallest enclosing quantum informational ball, to determine the information-theoretic security of the quantum communication by the maximally obtainable information of the eavesdropper.

*5.4.1 Phase-covariant Cloning*

In this section, we illustrate the quantum informational balls for the analyzed quantum cloners. The randomly chosen sent pure quantum states cloned by Eve's phase-covariant quantum cloner are denoted by $\rho_1, \rho_2, \rho_3$ and $\rho_4$. Using Delaunay tessellation, we compute the convex-hull of the cloned equatorial states $\rho_1, \rho_2, \rho_3$ and $\rho_4$. In Figure 30, we illustrate the convex-hull of cloned states in two- and three-dimensional Bloch-ball representations. The curved geometrical structure of quantum Delaunay tessellation is well observable.

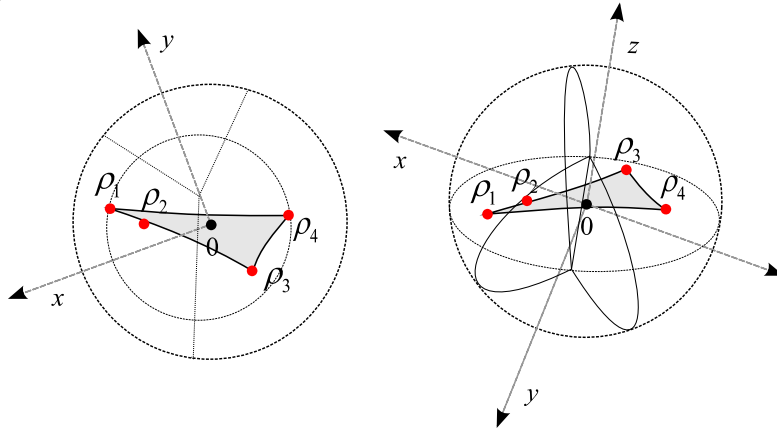

**Figure 30.** The convex hull of cloned mixed states.

From the convex set, we can compute the smallest enclosing quantum informational ball and its radius $r^*$. In Figure 31, we have illustrated the Euclidean smallest enclosing ball by the dashed circle, and the quantum relative entropy ball.

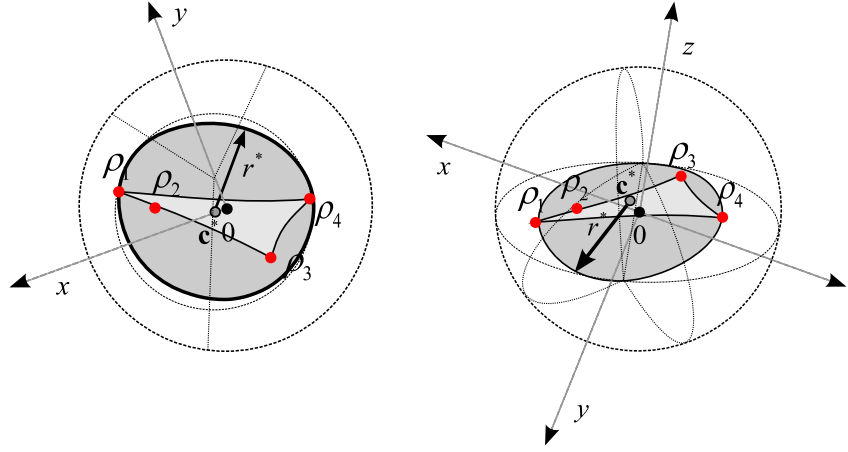

**Figure 31.** The smallest enclosing quantum informational balls.

From the smallest enclosing quantum informational ball, we can determine the radius $r^*$, which describes the information-theoretical impact of the eavesdropper cloning machine. The center of the smallest enclosing quantum informational ball is denoted by $\mathbf{c}^*$. We can analyze similarly the properties of universal quantum cloner.

*5.4.2 Universal Cloning*

In Figure 32(a), we have illustrated the dual Delaunay tessellation for the cloned states and the three-dimensional *convex hull* (light-grey) of cloned states $\rho_1, \rho_2, \rho_3, \rho_4, \rho_5$ and $\rho_6$. From the convex hull, we compute the smallest enclosing quantum informational ball. In Figure 32(b), we illustrate the smallest quantum informational ball and its radius $r^*$.

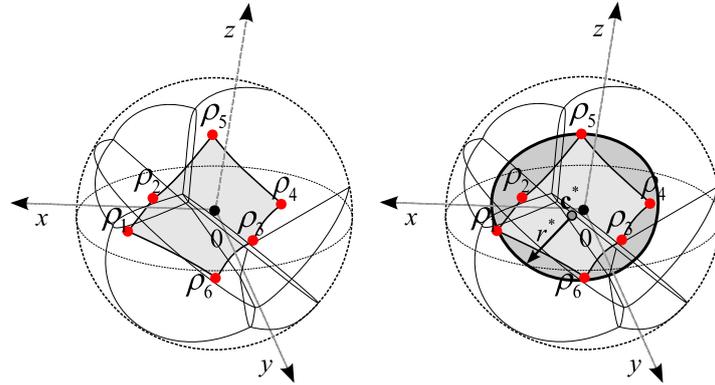

**Figure 32.** The convex hull (a) and the smallest quantum ball (b) of the cloned mixed states generated by a universal quantum cloner.

The radius of the smallest enclosing quantum informational ball of a phase-covariant cloner is greater for equatorial states than is that of a universal quantum cloner. On the other hand, the universal cloner shrinks all quantum states equally, while the phase-covariant cloner can be applied efficiently only for equatorial states.

## 6 Optimized Geometrical Algorithm

The quantum relative entropy-based algorithm at the $i$-th iteration gives an $\mathcal{O}(1+\sqrt{i})$-approximation of the real *circumcenter*, thus to get an $(1+\mathcal{E})$ approximation, our algorithm requires a time [13], [36-37]

$$\mathcal{O}\left(\frac{dn}{\varepsilon^2}\right) = \mathcal{O}\left(\frac{d}{\varepsilon^2}\frac{1}{\varepsilon}\right) = \mathcal{O}\left(\frac{d}{\varepsilon^3}\right), \tag{50}$$

by first sampling $n = \frac{1}{\varepsilon}$ points. Based on the computational complexity of the smallest enclosing ball, the $(1+\varepsilon)$ approximation of the fidelity of the eavesdropper cloning machine can be computed in a time

$$\mathcal{O}\left(\frac{d}{\varepsilon^2}\right). \tag{51}$$

In this section we improve our method to get a

$$\mathcal{O}\left(\frac{d}{\varepsilon}\right) \tag{52}$$

time, $(1+\varepsilon)$-approximation algorithm in quantum space [13], [36-37].

In the proposed algorithm, the optimal radius is between $r \leq r^* \leq r+\delta$, and the process terminates as $\delta \leq \varepsilon$, in at most $\mathcal{O}\left(\frac{1}{\varepsilon}\right)$ iterations. The main steps of the improved approximation algorithm are [13], [33-36]:

**Algorithm 2**.
1. *Select* a random center $\mathbf{c}_1$ from the set of quantum states $\mathcal{S}$
$$\mathbf{c}_1 \in \mathcal{S}$$
2. $r = \frac{1}{2}\max_i D_F(\mathbf{c}_1, \mathcal{S});$
3. $\delta = \frac{1}{2}\max_i D_F(\mathbf{c}_1, \mathcal{S});$
4. **for** $\left(i = 1, 2, \ldots, \left(\frac{1}{\delta}\right)\right)$
5.     **do**
6. $S = \arg\max_i D_F(c, \mathcal{S});$
7. Move $Ball(c, r)$ on the geodesic until it touches the *farthest* point $S$;

8. $s = \max_i D_F(c, S_i) - r$;
9.     **if**    $s \leq \frac{3\delta}{4}$ **then**
10.                        $\delta = \frac{3\delta}{4}$
11.     **else**
12.                        $r = r + \frac{\delta}{4}$;
13.                        $\delta = \frac{3\delta}{4}$;
14. **until** $\delta \leq \varepsilon$.

## 6.1 Rate of Convergence

In our experimental simulation, we have compared the core-set algorithm and our improved core-set algorithm to find the smallest enclosing quantum information ball. We have analyzed the approximation algorithms for 30 center updates and we have measured the quality of the approximation with respect to quantum relative entropy.

The results of our simulation are shown in Figure 33. The *x*-axis represents the number of center updates to find the center of the smallest enclosing quantum informational ball, the *y*-axis represents the quantum informational distance between the found center **c** and the optimal center $\mathbf{c}^*$.

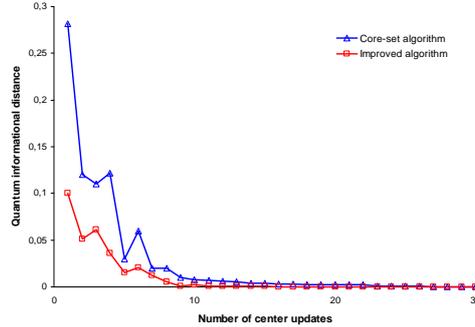

**Figure 33.** The rate of converge of approximation algorithms.

From the results, we can conclude that each algorithm finds the approximate center **c** to the optimal center $\mathbf{c}^*$ very fast. The quantum relative entropy-based approximation algorithms have a very accurate convergence of **c** towards $\mathbf{c}^*$, however the improved core-set algorithm converges faster with a smaller number of center updates.

## 7 Results

We have used the proposed computational geometric method to compute the radius of the smallest quantum informational ball, which measures the information obtained by the attacker.

In Figure 34 we show the results of our analysis for a collective attack. The radius of the eavesdropper's smallest enclosing quantum informational ball is described by Eq. (11). The effects of the disturbance caused by the eavesdropper is analyzed in the range of $[0, 0.5]$. The upper and lower bounds of radii of the eavesdropper's smallest quantum ball are shown as the function of the disturbance, using UCM and phase-covariant quantum cloners.

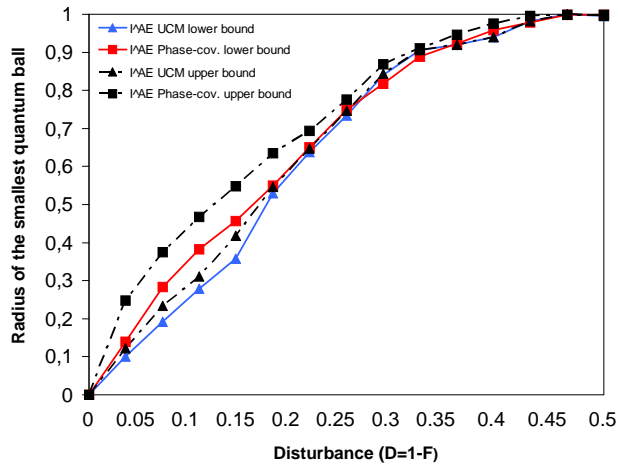

**Figure 34.** Results of information geometrical security analysis of DPS QKD protocol for collective attack. The eavesdropper's obtainable information is described by the radius of the smallest enclosing quantum informational ball. (lower bound - solid lines, upper bound - dashed lines)

We have used the mutual information analysis to show the security of the DPS QKD protocol against coherent attacks. The radius of the smallest enclosing informational ball, hence the maximal obtainable information of the eavesdropper, increases with the level of disturbance. However, in the tolerated range of the disturbance level of the DPS QKD protocol, the analyzed quantum cloners make no possible for an eavesdropper to realize a successful attack in practice. As it is well described by the radii of the smallest quantum informational balls, the UCM based attack allows Eve less information than the phase-covariant based attack, which result confirms the mutual information analysis of UCM and phase-covariant cloner based attacks [29, 30]. The results of the information-theoretic based analysis confirmed the fact, that coherent attack does not help to Eve to increase her information about the key.

# 8 Conclusions

The DPS QKD protocol has become more popular among quantum cryptographic protocols, since it offers higher key rates and easier implementation. The DPS protocol is tailored for practical applications, and represents a more applicable protocol than other discrete- and continuous-variable QKD protocols, which were invented by theorists. A bound for the DPS protocol's unconditional security is still missing. The differential phase-shift protocol has better rates and its practical realizations are much simpler. However the security of the DPS QKD protocol is still an open question, since it unconditional security has not been fully approved yet. The lower bounds on the security of DPS QKD scheme are currently an open question, and currently under research.

Our geometrical approach could be a very valuable tool to analyze the still open questions on the security of DPS QKD protocol. Our geometrical approach can be extended to all possible attacks against the DPS QKD protocol and the still unknown lower bounds can be discovered. In this paper, we analyzed the most general collective attack against the protocol by information geometrical methods, and for different attacker strategies. Our method could be a very valuable tool to analyze practical DPS QKD implementations, since for a large set of quantum states, an extremely efficient algorithm is required to compute the radius of the smallest quantum informational ball. The proposed information geometrical method analyses the information-theoretic security of the DPS QKD protocol, and it could offer a very useful practical algorithmical solution to solve the still open and unknown questions related to the information-theoretic security of quantum cryptographic protocols.

# Acknowledgement

The results discussed above are supported by the grant TAMOP-4.2.1/B-09/1/KMR-2010-0002, 4.2.2.B-10/1--2010-0009 and COST Action MP1006.